\documentclass[twocolumn,aps,prl,showpacs,amsmath,superscriptaddress]{revtex4}
\usepackage{amssymb}
\usepackage{mathrsfs}
\usepackage{graphicx}
\usepackage{float}

\begin{document}

\title{Topological superconductor to Anderson localization transition in one-dimensional incommensurate lattices}

\author{Xiaoming Cai}
\affiliation{Beijing National Laboratory for Condensed Matter
Physics, Institute of Physics, Chinese Academy of Sciences,
Beijing 100190, China}
\affiliation{State Key Laboratory of
Magnetic Resonance and Atomic and Molecular Physics, Wuhan
Institute of Physics and Mathematics, Chinese Academy of Sciences,
Wuhan 430071, China}

\author{Li-Jun Lang}
\affiliation{Beijing National Laboratory for Condensed Matter
Physics, Institute of Physics, Chinese Academy of Sciences,
Beijing 100190, China}

\author{Shu Chen}
\thanks{Corresponding author, schen@aphy.iphy.ac.cn}
\affiliation{Beijing National
Laboratory for Condensed Matter Physics, Institute of Physics,
Chinese Academy of Sciences, Beijing 100190, China}

\author{Yupeng Wang}
\affiliation{Beijing National Laboratory for Condensed Matter
Physics, Institute of Physics, Chinese Academy of Sciences,
Beijing 100190, China}

\date{\today}

\begin{abstract}
We study the competition of disorder and superconductivity for a
one-dimensional p-wave superconductor in incommensurate
potentials. With the increase in the strength of the
incommensurate potential, the system undergoes a transition from a
topological superconducting phase to a topologically trivial
localized phase. The phase boundary is determined both numerically
and analytically from various aspects and the topological
superconducting phase is characterized by the presence of Majorana
edge fermions in the system with open boundary conditions. We also
calculate the topological $Z_2$ invariant of the bulk system and
find it can be used to distinguish the different topological
phases even for a disordered system.
\end{abstract}


\maketitle

{\it Introduction.-} Topological superconductors (TSCs) have
attracted intense recent studies, as they are promising candidates
for the practical realization of Majorana fermions
\cite{Kitaev1,Lutchyn,Oreg,Ivanov,Stone,Kane,Beenakker}. Among
various proposals, the one-dimensional (1D) TSC in nanowires with
strong spin-orbit interactions and proximity-induced
superconductivity \cite{Lutchyn,Oreg} provides experimental
feasibility on the detection of Majorana fermions in hybrid
superconductor-semiconductor wires \cite{Mourik,Deng,Das}, which
has stimulated great enthusiasm in exploring physical properties
of topological superconductors. A key feature of a 1D TSC is the
emergence of edge Majorana fermions (MFs) at ends of the
superconducting (SC) wire as a result of bulk-boundary
correspondence. A prototype model unveiling topological features
of the 1D TSC is given by the effective spinless p-wave SC model
studied originally by Kitaev \cite{Kitaev1}.

As the TSC is protected by the particle-hole symmetry, the
topological phase is expected to be immune to perturbations of
weak disorder \cite{Potter}. Nevertheless, a strong disorder may
destroy the SC phase and induce a transition to the Anderson
insulator. Localization in 1D SC system in the presence of
disorder has been an active research field in the past decades
\cite{Altland,Brouwer2000,Motrunich,Gruzberg}. The theoretical
studies have unveiled that the particle-hole symmetry in the SC
system plays an important role in the problem of the Anderson
localization \cite{Altland}. Due to the existence of a finite SC
gap, the interplay of disorder and superconductivity leads to a
topological phase transition from topological SC phase to a
topologically trivial localized phase when the strength of
disorder increases over a critical value.

So far, most theoretical work for the Anderson localization in 1D
TSCs focuses on the random disorder
\cite{Brouwer2000,Motrunich,Gruzberg,Brouwer,Lobos}, disorder
produced by incommensurate potentials is concerned only very
recently \cite{Tezuka,Tezuka2}. While Ref. \cite{Tezuka} explores
the TS phase by tuning the chemical potential in 1D quantum wire
with spin-orbit interaction in proximity to a superconductor under
incommensurate modulation, we focus our study on the transition
from TS phase to Anderson localization purely induced by the
incommensurate potential for a 1D p-wave superconductor system. In
the absence of superconductivity, the localization transition
driven by the incommensurate potential occurs at a finite disorder
strength which can be exactly determined by a self-duality mapping
\cite{Aubry}, whereas an arbitrary weak random disorder induces
the Anderson localization in one dimension. The incommensurate
potential can now be engineered with ultracold atoms loaded in 1D
bichromatic optical lattices \cite{Roati}, opening the
experimental way to study the localization properties of
quasi-periodic systems. In this work, we shall study the interplay
of the incommensurate potential and topologically protected
superconductivity in the 1D p-wave SC model and determine the
phase boundary of TSC to localization transition exactly. The
tunability of the incommensurate potential \cite{Roati} provides a
potential way to experimentally study the controllable disorder
effect in TSCs realizable in cold atom systems \cite{Jiang}.

{\it Model of p-wave superconductor with incommensurate
potential.-} The 1D p-wave superconductor in the incommensurate
lattices is described by the following Hamiltonian:
\begin{eqnarray}
H = \sum_{i} [ (-t \hat{c}_{i}^{\dag } \hat{c}_{i+1}+ \Delta
\hat{c}_{i} \hat{c}_{i+1}+ H.c.) + V_{i} \hat{n}_{i} ],
\label{Ham}
\end{eqnarray}
where  $\hat{n}_{i}=\hat{c}^\dagger_i \hat{c}_i$ is the particle
number operator and $\hat{c}^\dagger_i$ ($\hat{c}_i$) the creation
(annihilation) operator of fermions. Here the nearest-neighbor
hopping amplitude $t$ and the p-wave pairing amplitude $\Delta$
are taken as real constants, whereas the incommensurate potential
\begin{equation}
V_{i}=V \cos(2\pi i \alpha)
\end{equation}
varies at each lattice site with $\alpha$ being an irrational
number and $V$ the strength of the incommensurate potential. The
model reduces to the Aubry-Andr\'{e} model when $\Delta=0$
\cite{Aubry}, while the Hamiltonian describes the Kitaev's p-wave
SC model for $\alpha=0$ \cite{Kitaev1}. For $\Delta=0$, the system
undergoes a delocalization to localization transition at $V=2t$.
On the other hand, the uniform p-wave SC system with $V_i=V$
undergoes a topological phase transition at $|V|=2t$ with a
topological nontrivial phase in the regime of $|V|<2t$
characterized by the presence of edge MFs \cite{Kitaev1}. In this
work, we shall study the interplay of the SC pairing $\Delta$ and
the incommensurate potential and then determine the phase diagram
of the system.

The Hamiltonian can be diagonalized by using the Bogoliubove-de
Gennes (BDG) transformation \cite{deGennes,Lieb}:
\begin{equation}
\eta _{n}^{\dag } = \sum_{i=1}^{L}[u _{n,i} \hat{c}_{i}^{\dag } +
v _{n, i} \hat{c}_{i}], \label{quasi}
\end{equation}
where $L$ is the number of lattice sites and $n =1, \cdots, L$.
Here $u _{n, i}$ and $v_{n, i}$ are chosen real. In terms of the
operators $\eta_n$ and $\eta_{n}^{\dagger}$, the diagonalized
Hamiltonian is written as $H=\sum_{n=1}^{L}\Lambda _{n}(\eta
_{n}^{\dag }\eta _{n}-\frac{1}{2})$ with $\Lambda _{n}$ being the
spectrum of the single quasi-particles. The spectrum as well as $u
_{n, i}$ and $v_{n, i}$ can be determined
by solving BDG equations:
\begin{eqnarray}
 \left(
\begin{array}{cc}
\hat{h} & \hat{\Delta} \\
-\hat{\Delta} & -\hat{h}%
\end{array}
\right)
 \left(
\begin{array}{c}
u _{n} \\
v _{n}%
\end{array}%
\right) =
\Lambda _{n} \left(
\begin{array}{c}
u_n \\
v_{n}%
\end{array}
\right), \label{BDG}
\end{eqnarray}
where $ \hat{h}_{ij} = -t (\delta_{j,i+1} + \delta_{j,i-1}) +
V_{i}  \delta_{ji}$, $\hat{\Delta}_{ij} = - \Delta
(\delta_{j,i+1}-\delta_{j,i-1})$,
$u_n^T=(u_{n,1},\cdots,u_{n,L})$ and
$v_n^T=(v_{n,1},\cdots,v_{n,L})$.
The symmetry of BDG equtions implies $ \eta_n (\Lambda _{n})=
\eta_n^{\dagger} (-\Lambda _{n})$. The ground state of the system
corresponds to the state with all negative quasi-particle energy
levels filled. If the quasi-particle energies are arranged in
ascending order, i.e., $\Lambda_i \leq \Lambda_{i+1}$, for
$\Lambda_i >0$,  the gap of the system is just given by $\Delta_g
= 2 \Lambda_1$. In the following calculation, we shall set $t=1$
as the energy unit.

{\it Transition from SC phase to disorder phase.-}
Numerically solving Eqs. (\ref{BDG}), we can get the whole
spectrum of quasi-particles. In Fig.1, we show the spectra for the
case of $\alpha = (\sqrt{5}-1)/2$ and $\Delta=0.5$ under periodic
boundary conditions (PBC). It is shown that there exists a regime
with obvious nonzero gaps when $V$ is smaller than a critical
value $V_c$. When $V$ exceeds the critical value, there is no an
obvious gap separating the negative and positive parts of spectra.
To see it more clearly, we show the variation of $\Delta_g$ versus
$V$ in the regime close to the transition point in Fig.2a. As
shown in the figure, the gap vanishes at about $V_c=3$ and the
system opens a very narrow gap in the regime of $V>V_c$. We
calculate the gap for systems with different $\Delta$ and find the
similar behavior: the gap reaches a minimum, which approaches zero
in the limit of $L \rightarrow \infty$, at the transition point
about $V_c = 2 + 2 \Delta$ and there exits a very narrow gap when
$V$ exceeds the transition point. For cases with different
irrational $\alpha$, we find similar phenomena and the transition
point does not depend on the specific choice of  $\alpha$
\cite{supplemental}.

\begin{figure}[tbp]
\includegraphics[width=1.0\linewidth]{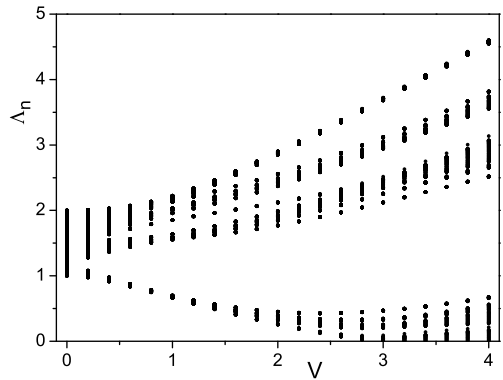}
\caption{Energy spectra of 1D p-wave superconductors with
$\alpha=(\sqrt{5}-1)/2$, $\Delta=0.5$ and $L=500$ under PBC. }
\end{figure}
\begin{figure}[tbp]
\includegraphics[width=1.1\linewidth]{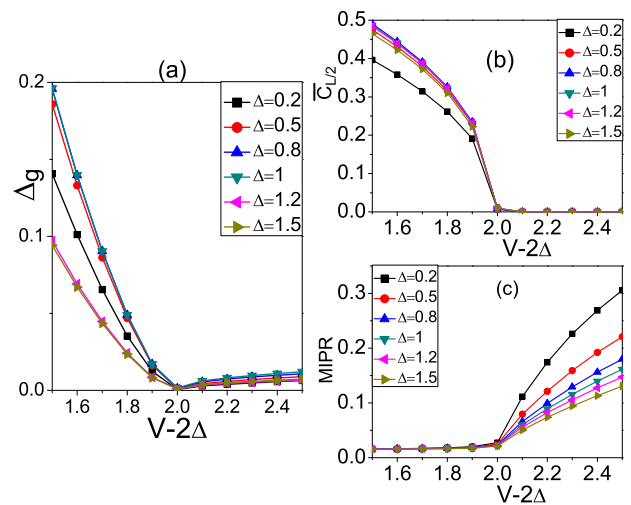}
\caption{(a). The energy gap $\Delta_g$ versus $V-2\Delta$, (b).
the average correlation function $\overline{C}_{L/2}$ versus
$V-2\Delta$, (c). the MIPR versus $V-2\Delta$ for the system with
$\alpha = (\sqrt{5}-1)/2$ and $L=500$. }
\end{figure}

Observing that the p-wave fermion model corresponds to the
transverse XY model with a randomly (irrationally) modulated
transverse field \cite{Fisher,Young}:
$\hat{H}= - \sum_{i}[ J_x\sigma_{i}^{x}\sigma_{i+1}^{x} + J_y
\sigma_{i}^{y} \sigma_{i+1}^{y} ] + \sum_{i} h_i \sigma_{i}^{z}$,
with the identification of $J_x = (t + \Delta)/2$, $J_y=
(t-\Delta)/2$ and $h_i=-V_i/2$, we can identify the phase
transition by calculating the correlation function $C_{ij}=\langle
\sigma_i^{x} \sigma_j^{x} \rangle$. In the language of quantum
spin model, the ferromagnetic phase is characterized by the
long-range order of the correlation function $
\langle\sigma^x_i\sigma^x_j\rangle_{|i-j|\longrightarrow\infty}=A
$ with $A$ being a nonzero positive number. In the original
fermion representation, $\sigma_i^{x}=(\hat{c}^\dagger_i +
\hat{c}_i)\mathrm{exp}(-i\pi\sum_{j=1}^{i-1}\hat{c}^\dagger_j
\hat{c}_j)$ takes a nonlocal form including a string product of
fermion operators, and the correlation function $C_{ij}=\langle
(\hat{c}_{i}^{\dag } + \hat{c}_{i}) \exp(-i\pi \sum_{l=i}^{j}
\hat{n}_l )(\hat{c}_{j}^{\dag } + \hat{c}_{j})\rangle$. In the
presence of the disordered potential, the correlation function
$C_{ij}$ will oscillate and we define the average correlation
function $\overline{C}_{r}=  \sum_{i} C_{i,i+r}/L$. Then for a
large system under PBC, the value of $\overline{C}_{L/2}$ can be
used to distinguish the SC phase and the localized phase. The
correlation function $C_{ij}$ can be calculated by the exact
numerical method described in Ref.\cite{Young}. In Fig.2b we show
the relation between $\overline{C}_{L/2}$ and $V$ for systems with
different $\Delta$. Without the disordered potential, the
correlation function $\overline{C}_{L/2}$ is a positive number and
increases as $\Delta$ increases for $0<\Delta<1$, gets its largest
value $\overline{C}_{L/2}=1$ at $\Delta=1$, then decreases for
$\Delta>1$. As the strength of $V$ increases, $\overline{C}_{L/2}$
decreases monotonically and approaches zero when $V- 2 \Delta$ is
about $2$. When $V > 2 + 2 \Delta$, the system loses the long
range order of correlation function and the system is driven into
the Anderson localized phase.

To characterize the localization transition, we define the
quantity of the inverse participation ratio (IPR) as $P_n =
\sum_{i=1}^{L} (u_{n,i}^4 + v_{n,i}^4)$, where $u_{n,j}$ and $
v_{n,j}$ are the solution to BDG equations and fulfil the
normalization condition $ \sum_i (u_{n,i}^2 + v_{n,i}^2)=1$. The
above definition can be viewed as an extension of IPR for the case
with $\Delta=0$ \cite{IPR,IPR2}. For an extended state, $P_n
\rightarrow 1/L$ and the IPR tends to zero for large $L$, whereas
the IPR tends to a finite number for a localized state. Therefore,
IPR can be taken as a criterion to distinguish the extended states
from the localized ones. Since the ground state is composed of
states with all negative quasi-particle energy levels filled, we
define the mean inverse participation ratio (MIPR)  as
$\text{MIPR} = \sum_{n=1}^{L} P_n / L$ to characterize the
localization of the ground state. As shown in Fig.2c, the MIPR
increases monotonically with the increase of $V$. At $V = 2 + 2
\Delta$, the MIPR has a sudden increase which characterizes a
localization transition. As a comparison, we note that the
localization transition does not occur for the commensurate
potential system with a rational $\alpha$ \cite{Lang}, for which
the wave functions of a periodic system take the Bloch's form and
are extended for arbitrary $V$.
\begin{figure}[tbp]
\includegraphics[width=1.0 \linewidth]{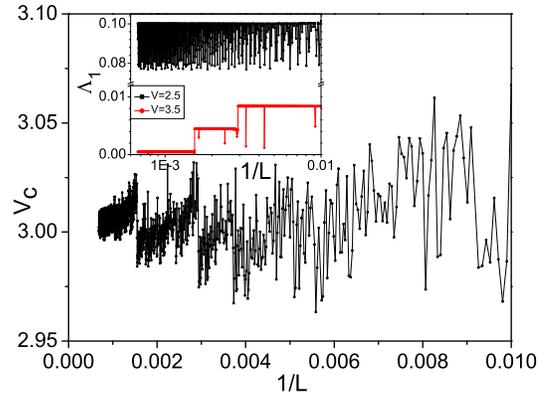}
\caption{ The finite size analysis of the transition point, i.e.,
$V_c(L)$ vs $1/L$. Inset: $\Lambda_1(L) = \Delta_g(L)/2$ vs $1/L$
in the regime of $V<V_c$ and $V>V_c$.}
\end{figure}

We then make finite size analysis by calculating the transition
points for systems with different sizes. As shown in Fig.3, the
value of transition points $V_c(L)$ for systems with $\Delta=0.5$
oscillates around $3.0$. Defining $V_{avc}=
\sum_{L=L_{min}}^{L_{max}} V_c(L)/(L_{max}-L_{min})$, we calculate
the average of $V_c(L)$ for different $L_{max}$ and $L_{min}$ and
find that $V_{avc}$ is about $3.0040 \pm 0.0005$ being very close
to $3$. The change of the gap size at $V=2.5$ and $V=3.5$ is shown
in the inset of Fig.3, which indicates that the gap is finite in
the regime of $V<V_c$ whereas the narrow gap in the regime of
$V>V_c$ approaches zero in the large $L$ limit. We also check
systems with different $\Delta$ and find similar behaviors, i.e.,
$V_c(L)-2\Delta$ oscillates with $L$ and approaches to $2.0$ in
the large $L$ limit.

Next we make analytical derivation of the critical value $V_c$ in
the large $L$ limit \cite{supplemental}. Rewriting the Hamiltonian
(\ref{Ham}) as the form of $H= \sum_{ij}[\hat{c}_{i}^{\dagger
}A_{ij}\hat{c}_{j}+\frac{1}{2}(\hat{c}_{i}^{\dagger
}B_{ij}\hat{c}_{j}^{\dagger }+h.c.)]$, where $A$ is a Hermitian
matrix and $B$ is an antisymmetric matrix, we can obtain the
excitation spectrum $\Lambda _{n}$ by solving the secular equation
$ \mathrm{det}[(A+B)(A-B)-\Lambda _{n}^{2}]=0$
\cite{Lieb,supplemental}. Since the excitation gap approaches zero
at the phase transition point, $V_c$ can be determined by the
condition of  $ \mathrm{det}[(A-B)(A+B)]=0$. By using the relation
$ \mathrm{det}(A-B)=\mathrm{det}(A-B)^{T}=\mathrm{det}(A+B)$, we
can determine $V_c$ by  $\mathrm{det}(A-B)=0$, which leads to the
constraint condition
\begin{equation}
\prod_{i=1}^{L}\mathrm{cos}(2\pi \alpha i)=\left( \frac{\Delta +t}{V}
\right) ^{L}
\end{equation}
in the limit of $L \rightarrow \infty$. Taking logarithm of the
above equation and replacing the summation by integral, we can get
$ V_{c}= 2(\Delta +t)e^{i2\pi n/L}$ with $n$ being the integer.
For the real solution of $V_{c}$, we have $ \left\vert
V_{c}\right\vert =2(\Delta +t) $, which is consistent with our
numerical result.

{\it Topological features of the topological SC phase.-}
To characterize the topological properties of the SC phase, we
seek the zero-mode solution of the system under open boundary
conditions (OBC). As shown in Fig.4a, we plot the quasi-particle
spectra of BDG equations under OBC. In comparison with the spectra
under PBC, an obvious feature is the the presence of the zero mode
solution in the gap regime.
The enlarged $\Lambda_1$ is shown in the inset of Fig.4a, which
indicates a sudden increase in $\Lambda_1$ for $V>3$. Here
the zero mode solution corresponds to the Majorana edge state with MFs
localized at ends of 1D wires. To see it clearly,
we introduce the Majorana operators $\gamma_i^A =
\hat{c}_{i}^{\dag }+ \hat{c}_{i}$ and $\gamma_i^B = (\hat{c}_{i}-
\hat{c}_{i}^{\dag })/i$, which fulfill the relations
$(\gamma_i^\alpha)^{\dagger} = \gamma_i^\alpha$ and
anticommutation relations $\{\gamma_i^\alpha, \gamma_i^\beta\} = 2
\delta_{ij} \delta_{\alpha \beta}$ with $\alpha$ and $\beta$
taking $A$ or $B$, and rewrite the quasi-particle operators as
\begin{eqnarray}
\eta _{n}^{\dag } = \frac{1}{2}\sum_{i=1}^{L}[ \phi _{n, i}
\gamma_i^A  - i \psi _{n,i} \gamma_i^B ],
\end{eqnarray}
where $\phi _{n, i}=(u_{n, i}+ v_{n, i})$ and $ \psi _{n,
i}=(u_{n, i} - v_{n, i})$. Typical distributions of $\phi_i$ and
$\psi_i$ for the lowest excitation solution of $\Lambda_1$ are
shown in Fig.4b and Fig.4c. When $V<V_c$, $\phi_i$ ($\psi_i$) is
located at the left (right) end and decays very quickly away from
the left (right) edge. As $V$ deviates farther from the
transition point $V_c$, the edge mode decays more
quickly. Since there is no overlap for the
amplitudes of $\gamma_i^A$ and $\gamma_i^B$, the zero mode fermion
splits into two spatially separated MFs. On the contrary, distributions of $\phi_i$
and $\psi_i$ for the lowest excitation mode in the regime of
$V>V_c$, for example $V=3.5$, overlap together
and locate inside of the bulk as a
result of Anderson localization. Consequently, the corresponding
quasiparticle is a localized fermion which can not be split into
two independent MFs. Therefore, the transition from
TSCs to Anderson localizations can be also judged by the presence
or absence of edge MFs in different parameter
regimes of the system with OBC.
\begin{figure}
\includegraphics[width=0.9\linewidth]{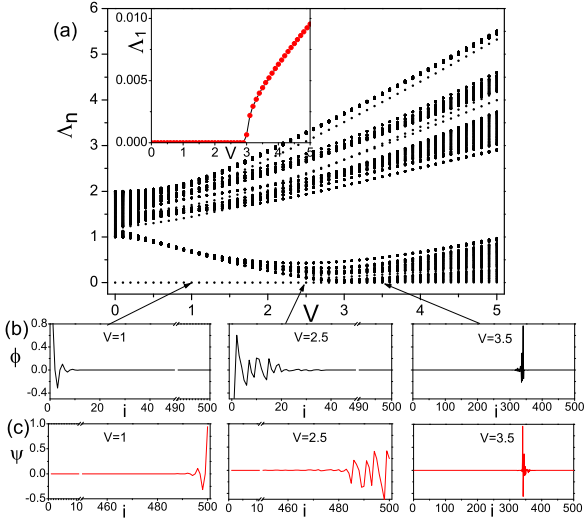}
\caption{ (a) Energy spectra of 1D p-wave superconductors with
$\alpha=(\sqrt{5}-1)/2$, $\Delta=0.5$ and $L=500$ under OBC. The
spatial distributions of $\phi_i$ (b) and $\psi_i$ (c) for the
lowest excitation with various $V$. }
\end{figure}


\begin{figure}[tbp]
\includegraphics[width=1.1\linewidth]{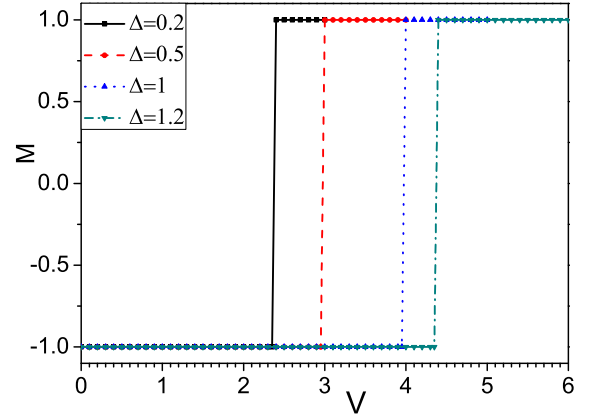}
\caption{$Z_2$ topological invariant vs $V$ for systems with
$\alpha=(\sqrt{5}-1)/2$, $L=500$ and various $\Delta$. }
\end{figure}

{\it $Z_2$ topological invariant.-} The existence of Majorana edge
states is attributed to the nontrivial topological nature of the
bulk superconductor, which can be characterized by a $Z_2$
topological invariant \cite{Kitaev1}. In terms of Majorana
operators, the Hamiltonian (1) can be represented as
$H=\frac{i}{4}\sum_{l,m=1}^{2L}A_{lm}\gamma_l\gamma_m$ with $\quad
A^*_{lm}=A_{lm}=-A_{ml}$, where $L$ is the number of lattice
sites, $A$ is a skew-symmetric matrix, $\gamma_l$ is defined as
$\gamma_{2j-1} = \gamma^A_j,\gamma_{2j}= \gamma^B_j$ and
$\{\gamma_l,\gamma_m\}=2\delta_{lm}$. The nonzero matrix elements
are given by $ A_{2j-1,2j}= -A_{2j,2j-1}= V \cos(2\pi j \alpha)$,
$A_{2j-1,2j+2}=-A_{2j+2,2j-1}=\Delta-1$ and
$A_{2j,2j+1}=-A_{2j+1,2j}=1 + \Delta $ for $j=1,\cdots,L$ with the
boundary condition of $L+1=1$. For a skew-symmetric matrix $A$,
the Pfaffian is defined as
$\mathrm{Pf}(A)=\frac{1}{2^LL!}\sum_{\tau\in
S_{2L}}\mathrm{sgn}(\tau)A_{\tau(1),\tau(2)}\cdot\cdot\cdot
A_{\tau(2L-1),\tau(2L)}$, where $S_{2L}$ is the set of
permutations on $2L$ elements and $\mathrm{sgn}(\tau)$ is the sign
of permutation. With Pfaffian of a system, the $Z_2$ topological
invariant is defined as $M=\mathrm{sgn}(\mathrm{Pf}(A))$. As shown
in Fig.5, the $Z_2$ topologically non-trivial phase is
characterized $M =-1$, whereas the $Z_2$ topologically trivial
phase corresponds to $M = 1$. For the system with $V<2+2\Delta$,
the $Z_2$ number $M=-1$ and the system is in the topologically
non-trivial phase, while for the system with $V>2+2\Delta$, the
$Z_2$ number $M=1$ and the system is in the topologically trivial
phase. As the strength of $V$ increases, a topological phase
transition happens.

{\it Summary.-} In summary, we study the effect of disorder
produced by the incommensurate potential in 1D p-wave
superconductors which support a topological SC phase with Majorana
edge states. Increasing the strength of disorder destroys the
topological SC phase and drives the system into a Anderson
localized state. The phase transition driven by the disorder is
identified by analyzing the change of gap, the long-range order of
the correlation function of nonlocal operators and the IPR which
characterizes the spacial localization of wavefunctions. The
transition point is exactly determined both numerically and
analytically. A $Z_2$ topological invariant is also used to
identify the transition from the topological SC phase, which has
emergent Majorana edge states for the system with OBC, to the
topologically trivial localized state.

\begin{acknowledgments}
This work has been supported by National Program for Basic
Research of MOST, NSF of China under Grants No.11174360 and
No.11121063, and 973 grant.
\end{acknowledgments}

{\it Note added.} During the preparation of this manuscript we
became aware of a preprint on similar topics \cite{DeGottardi}.


\onecolumngrid
\newpage
\section{Supplemental material for ``Topological superconductor to Anderson localization transition in one-dimensional incommensurate lattices"}

\subsection{I. Examples for different irrational $\alpha$}
\begin{figure}[bp]
\includegraphics[width=0.6\linewidth]{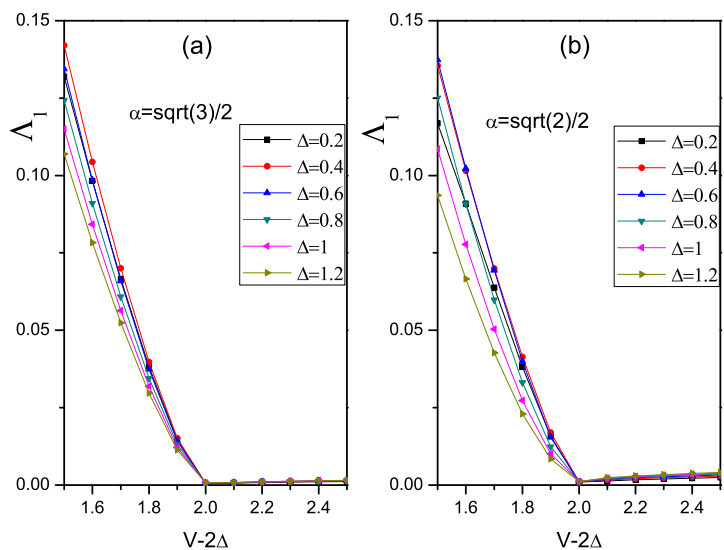}
\caption{The lowest excitation energy, $\Lambda_1$, versus
$V-2\Delta$ for the system with $L=500$, $\alpha=\sqrt{3}/2$ and
$\alpha=\sqrt{2}/2$ . } \label{fig7}
\end{figure}

In the main text, we discuss the topological superconductor to
Anderson localization transition in one-dimensional incommensurate
lattices by considering the case with $\alpha=(\sqrt{5}-1)/2$
(inverse golden ratio). The conclusions obtained in the main text
do not depend on the choice of the inverse golden ratio. In the
supplemental material, we give two more examples for the
incommensurate potential with some other irrational values. To
give concrete examples, we show the variation of $\Lambda_1=
\Delta_g/2$ versus $V$ in the regime close to the transition point
for $\alpha=\sqrt{3}/2$ and $\alpha=\sqrt{2}/2$ in
Fig.{\ref{fig7}}. As shown in the figure, for both cases with
$\alpha=\sqrt{3}/2$ and $\alpha=\sqrt{2}/2$, the gap vanishes at
about $V_c=2 + 2 \Delta$ for different $\Delta$. In comparison
with Fig.2(a) in the main text, we can see that the transition
point does not depend on the choice of  $\alpha=(\sqrt{5}-1)/2$ as
long as $\alpha$ being the irrational number. For the case of
$\Delta=0$, it is known that the transition point from extended
states to localized states at $V_c=2$ does not depend on the
special choice of inverse golden ratio.
\begin{figure}[tbp]
\includegraphics[width=0.6\linewidth]{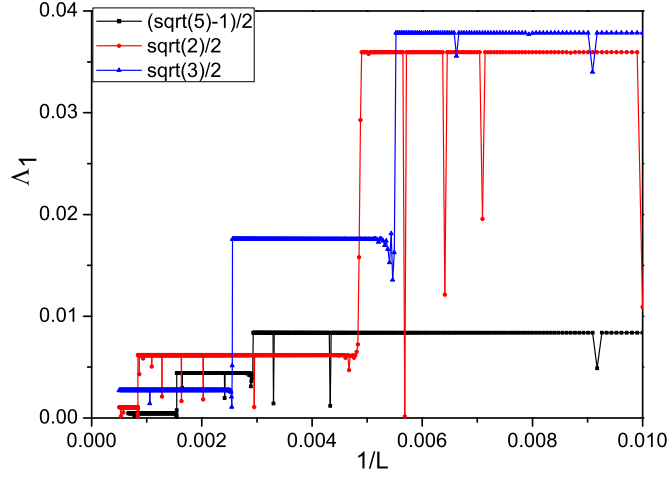}
\caption{The finite size analysis of $\Lambda_1$ for the system
with $V=3.5$, $\Delta=0.5$ and different $\alpha$. } \label{fig8}
\end{figure}

Next we show finite size analysis of the lowest excitation energy,
$\Lambda_1$, in the regime of $V > V_c$ for cases with
$\alpha=\sqrt{3}/2$ and $\alpha=\sqrt{2}/2$. For systems with
$\Delta=0.5$, the change of $\Lambda_1$ at $V=3.5$ is shown in
Fig.{\ref{fig8}}, which indicates that the narrow gap in the
regime of $V>V_c$ approaches zero  in the large $L$ limit in a
similar way as the case of $\alpha=(\sqrt{5}-1)/2$.

\subsection{II. Details for the derivation of phase transition point}

Under periodic boundary conditions, $\hat{c}_{L+1}^{\dagger
}=\hat{c}_{1}^{\dagger}$, the Hamiltonian (1) in the main text can
be represented as
\begin{eqnarray}
H &=&\sum_{i=1}^{L}[- t \hat{c}_{i}^{\dagger }
\hat{c}_{i+1}+\Delta
\hat{c}_{i} \hat{c}_{i+1} +h.c.+V_{i} \hat{n}_i] \nonumber \\
&=& \sum_{ij}[\hat{c}_{i}^{\dagger
}A_{ij}\hat{c}_{j}+\frac{1}{2}(\hat{c}_{i}^{\dagger
}B_{ij}\hat{c}_{j}^{\dagger }+h.c.)],
\end{eqnarray}%
where $A,B$ are $L\times L$ ($L$ is the number of lattice sites) matrices as
\begin{equation*}
A=\left(
\begin{array}{ccccc}
V_{1} & -t &  & \cdots & -t \\
-t & V_{2} & -t &  &  \\
& -t & V_{3} &  &  \\
\vdots &  &  & \ddots & -t \\
-t &  &  & -t & V_{L}%
\end{array}%
\right) ,B=\left(
\begin{array}{ccccc}
0 & -\Delta &  & \cdots & \Delta \\
\Delta & 0 & -\Delta &  &  \\
& \Delta & 0 &  &  \\
\vdots &  &  & \ddots & -\Delta \\
-\Delta &  &  & \Delta & 0%
\end{array}%
\right)
\end{equation*}%
with $V_{i}=V\mathrm{cos}(2\pi \alpha i )$.

The excitation spectrum $\Lambda _{n}$ is a solution of the secular
equation, \cite{Lieb}
\begin{equation*}
\mathrm{det}[(A+B)(A-B)-\Lambda _{n}^{2}]=0.
\end{equation*}%
For the system undergoing a topological quantum phase transition,
the excitation gap has to be closed at the phase transition point.
So the above
equation must have a solution $\Lambda _{n}=0$ at the critical point, i.e., $%
\mathrm{det}[(A-B)(A+B)]=0$. Notice that%
\begin{equation*}
\mathrm{det}(A-B)=\mathrm{det}(A-B)^{T}=\mathrm{det}(A+B),
\end{equation*}%
so the phase transition point can be determined by the following requirement%
\begin{equation}
\mathrm{det}(A-B)=0.  \label{cond}
\end{equation}

Now we need calculate the determinant of\ matrix $A-B$, that is
\begin{eqnarray}
&&\mathrm{det}(A-B)  \notag \\
&=&-(t-\Delta)^{L}-(t+\Delta)^{L}+V^{L}\prod_{i=1}^{L}\mathrm{cos}(2\pi
\alpha i )  \notag \\
&&+V^{L-2}(\Delta ^{2}-t^{2})\sum\limits_{j=1}^{L}\prod_{\substack{ i=1  \\ %
i\neq j,j+1}}^{L}\mathrm{cos}(2\pi \alpha i)+V^{L-4}(\Delta
^{2}-t^{2})^{2}\sum\limits_{j_{1}=1}^{L}\sum\limits_{j_{2}=j_{1}+2}^{L}\prod
_{\substack{ i=1  \\ i\neq j_{1},j_{1}+1,  \\ j_{2},j_{2}+1}}^{L}\mathrm{cos}%
(2\pi \alpha i)  \notag \\
&&+\cdots +V^{L-2n}(\Delta ^{2}-t^{2})^{n}\sum\limits_{\substack{ %
\{j_{s}=j_{s-1}+2,  \\ s=1,\cdots ,n\}}}^{L}\prod_{\substack{ i=1
\\ i\neq j_{s},j_{s}+1  \\ (s=1,\cdots ,n)}}^{L}\mathrm{cos}(2\pi
\alpha i )+\cdots .  \label{sum}
\end{eqnarray}
We note that the inverse golden ratio $(\sqrt{5}-1)/2$ can be
approached by a set of rational numbers $F_{n-1}/F_{n}$ when
$n\rightarrow \infty $, where $F_{n}$ is the Fibonacci sequence
given by $F_{n}=F_{n-1}+F_{n-2}$ with $F_1=F_0=1$. As an
irrational number can always be approximated by some rational
number, without loss of generality, we take $\alpha =p/q$ ($p,q$
are co-prime integers) with $q \rightarrow \infty$ to approach a
given irrational number. For a finite size system, it is
physically impossible to distinguish between an irrational
$\alpha$ and its rational approximation $p/q$ as long as $q \geq
L$. To make progress, we shall evaluate the determinant of the
$L\times L$ matrix $A-B$, by taking  $\alpha=p/q$ with $L=q$ (for
the inverse golden ratio $L=F_n$). Under the condition of
$\alpha=p/L$ ($p$ is co-prime to $L$), Eq. (\ref{sum}) is greatly
simplified as the summation terms including $\mathrm{cos}(2\pi
\alpha i )$ vanish for each $n$ except $n=0$ and $L/2$. To see it
clearly, one can check it order by order with the increase of $L$.
For examples, for $L=3$, the term except $n=0$ is
$\sum\limits_{i=1}^{3}\mathrm{cos}(\frac{2 \pi p}{3}%
i)=0$; for $L=4$, the term except $n=0$ and $2$ is
$\sum\limits_{i=1}^{4}\mathrm{cos}(\frac{2\pi p }{4}%
i )\mathrm{cos}[\frac{2\pi p }{4}(i+1) ]=0$; for $L=5$, the terms
except $n=0$ are
$ \sum\limits_{i=1}^{5}\mathrm{cos}(\frac{2 \pi p }{5}%
i)\mathrm{cos}[\frac{2\pi p }{5}(i+1)]\mathrm{cos}[\frac{2 \pi p
}{5}(i+2) ] =0$ and
$\sum\limits_{i=1}^{5}\mathrm{cos}(\frac{2 \pi p }{5}%
i) =0$; and for $L=6$, the terms except $n=0$ and $3$ are
$\sum\limits_{i=1}^{6}\mathrm{cos}(\frac{2\pi p }{6}%
i)\mathrm{cos}[\frac{2 \pi p }{6}(i+1)]\mathrm{cos}[\frac{2\pi
p}{6}(i+2) ]\mathrm{cos}[\frac{2\pi
p}{6}(i+3) ] =0$ and $\sum\limits_{i=1}^{6}\mathrm{cos}(\frac{2\pi p}{%
6}i )\mathrm{cos}[\frac{2\pi p}{6}(i+1)]+\sum\limits_{i=1}^{3}%
\mathrm{cos}(\frac{2\pi p}{6}i )\mathrm{cos}[\frac{2\pi p}{6}%
(i+3)] =0$. For larger $L$, one can also find that all the terms
vanish except $n=0$ and $L/2$, only leaving
$V^{L}\prod_{i=1}^{L}\mathrm{cos}(2\pi \alpha i)$ and $(\Delta
^{2}-t^{2})^{L/2}$, i.e.%
\begin{equation}
\det (A-B)=\left\{
\begin{array}{l}
\prod_{i=1}^{L}V\mathrm{cos}(2\pi \alpha i
)-(t-\Delta)^{L}-(\Delta
+t)^{L},\ \text{for odd }L \\
\prod_{i=1}^{L}V\mathrm{cos}(2\pi \alpha i )-(t-\Delta
)^{L}-(\Delta
+t)^{L}+(\Delta ^{2}-t^{2})^{L/2},\ \text{for even }L .%
\end{array}%
\right.   \label{det}
\end{equation}

From the above discussion, we can infer that Eq. (\ref{det}) is
still valid for irrational $\alpha $'s with $L=q\rightarrow \infty
$. In the limit of $L\rightarrow \infty $, the condition to find
the critical point, Eq. (\ref{cond}), can be rewritten as
\begin{equation*}
\left\{
\begin{array}{l}
\prod_{i=1}^{L}\frac{V}{\Delta +t}\mathrm{cos}(2\pi \alpha i )-(\frac{%
t-\Delta }{t+\Delta})^{L} - 1 =0,\ ~~~ \text{for odd }L \\
\prod_{i=1}^{L}\frac{V}{\Delta +t}\mathrm{cos}(2\pi \alpha i )-(\frac{%
t-\Delta }{t+\Delta })^{L}+(\frac{\Delta-t }{\Delta +
t})^{L/2}-1=0,\ ~~~ \text{for
even }L .%
\end{array}%
\right.
\end{equation*}%
Without loss of generality, we suppose that $\Delta ,t>0$, and the above
equations reduce to
\begin{equation}
\prod_{i=1}^{L}\mathrm{cos}(2\pi \alpha i)=\left( \frac{\Delta +t}{V%
}\right) ^{L}
\end{equation}%
in the limit of $L\rightarrow \infty $. The above equation is equivalent to%
\begin{eqnarray*}
\sum_{i=1}^{L}\ln \cos (2\pi \alpha i ) &=&\ln \left( \frac{\Delta +t%
}{V}\right) ^{L}+i2\pi n,\ n\in \text{Integer}
\end{eqnarray*}
or
\begin{eqnarray}
\frac{1}{L}\sum_{i=1}^{L}\ln \cos (2\pi \alpha i) &=&\ln \left(
\frac{\Delta +t}{V}\right) +\frac{i2\pi n}{L},\ n\in
\text{Integer}.
\end{eqnarray}%
For the irrational $\alpha$, we can replace the summation on the
left hand side of the above equation by integral
in the limit $L\rightarrow \infty $, that is%
\begin{eqnarray*}
\frac{1}{L}\sum_{i=1}^{L}\ln \mathrm{\cos (}2\pi \alpha i)
&\rightarrow &\int_{0}^{1}\ln \mathrm{\cos }(2\pi \alpha Lx)dx \\
&=&\frac{1}{2\pi \alpha L}\int_{0}^{2\pi \alpha L}\ln
\mathrm{cos}(x)dx = -\frac{1}{2\pi \alpha L}\mathcal{L}(2\pi
\alpha L),
\end{eqnarray*}%
where $\mathcal{L}(x)$ is the Lobachevskiy's function
\cite{Gradshteyn} defined as
\begin{eqnarray*}
\mathcal{L}(x) &=&-\int_{0}^{x}\ln \mathrm{\cos }\text{(}t)dt \\
&=&x\ln 2-\frac{1}{2}\sum_{k=1}^{\infty }(-1)^{k-1}\frac{\sin (2kx)}{k^{2}}.
\end{eqnarray*}%
Therefore, we get
\begin{eqnarray*}
\lim_{L \rightarrow \infty} \frac{1}{L}\sum_{i=1}^{L}\ln
\mathrm{\cos (}2\pi \alpha i )= -\ln 2.
\end{eqnarray*}%
So the critical potential strength $V_{c}$ is determined by $-\ln
2 =\ln \left( \frac{\Delta +t}{V_{c}}\right) +\frac{i2\pi n}{L} $,
i.e.,
\begin{eqnarray*}
V_{c} &=& 2(\Delta +t)e^{i2\pi n/L}, ~~~ \ n\in \text{Integer}.
\end{eqnarray*}%
For the real solution of $V_{c}$, we have
\begin{equation}
\left\vert V_{c}\right\vert =2(\Delta +t),
\end{equation}%
which gives the topological quantum phase transition point from
topological superconductor to Anderson localization. The above
derivation is not limited to the case of $\Delta,t>0$. For general
cases, similar derivation can be directly followed and $V_{c}$ is
given by
\begin{equation}
\left\vert V_{c}\right\vert =2(\left\vert \Delta \right\vert +
\left\vert t \right\vert).
\end{equation}%

\end{document}